\documentclass[twocolumn]{aastex62}
\usepackage{natbib}
\usepackage{color,subfigure,lineno} 

\newcommand{\etal}{et al.}
\newcommand{\hbeta}{H{$\beta$}}
\newcommand{\halpha}{H{$\alpha$}}

\newcommand{\CIV}{C\,{\sevenrm IV}}

\newcommand{\CIII}{C\,{\sevenrm III]}}

\newcommand{\AlIII}{Al\,{\sevenrm III}}

\newcommand{\SiIII}{Si\,{\sevenrm III]}}

\def\MgII{Mg\,{\sc ii}}

   \font\sevenrm=cmr7 scaled 1000
\newcommand{\comments}[1]{}

\shorttitle{SDSS-RM: High-$z$ RM lags}
\shortauthors{Shen et al.}

\begin{document}

\title{The Sloan Digital Sky Survey Reverberation Mapping Project: Improving Lag Detection with an Extended Multi-Year Baseline}


\author[0000-0002-6893-3742]{Yue Shen}
\altaffiliation{Alfred P. Sloan Research Fellow}
\affiliation{Department of Astronomy, University of Illinois at Urbana-Champaign, Urbana, IL 61801, USA}
\affiliation{National Center for Supercomputing Applications, University of Illinois at Urbana-Champaign, Urbana, IL 61801, USA}

\author{C.~J.~Grier}
\affiliation{Department of Astronomy and Astrophysics, Eberly College of Science, The Pennsylvania State University, 525 Davey Laboratory, University Park, PA 16802}
\affiliation{Institute for Gravitation \& the Cosmos, The Pennsylvania State University, University Park, PA 16802}
\affiliation{Steward Observatory, The University of Arizona, 933 North Cherry Avenue, Tucson, AZ 85721, USA} 

\author{Keith~Horne}
\affiliation{SUPA Physics and Astronomy, University of St. Andrews, Fife, KY16 9SS, Scotland, UK} 

\author{W.~N.~Brandt}
\affiliation{Department of Astronomy and Astrophysics, Eberly College of Science, The Pennsylvania State University, 525 Davey Laboratory, University Park, PA 16802}
\affiliation{Institute for Gravitation \& the Cosmos, The Pennsylvania State University, University Park, PA 16802}
\affiliation{Department of Physics, The Pennsylvania State University, University Park, PA 16802, USA}

\author{J.~R.~Trump}
\affiliation{Department of Physics, University of Connecticut, 2152 Hillside Road, Unit 3046, Storrs, CT 06269, USA}

\author{P.~B.~Hall}
\affiliation{Department of Physics and Astronomy, York University, Toronto, ON M3J 1P3, Canada}

\author{K.~Kinemuchi} 
\affiliation{Apache Point Observatory and New Mexico State University, P.O. Box 59, Sunspot, NM, 88349-0059, USA}

\author{David~Starkey} 
\affiliation{SUPA Physics and Astronomy, University of St. Andrews, Fife, KY16 9SS, Scotland, UK} 
\affiliation{Department of Astronomy, University of Illinois at Urbana-Champaign, Urbana, IL 61801, USA}

\author{D.~P.~Schneider}
\affiliation{Department of Astronomy and Astrophysics, Eberly College of Science, The Pennsylvania State University, 525 Davey Laboratory, University Park, PA 16802}
\affiliation{Institute for Gravitation \& the Cosmos, The Pennsylvania State University, University Park, PA 16802}

\author{Luis~C.~Ho}
\affiliation{Kavli Institute for Astronomy and Astrophysics, Peking University, Beijing 100871, China} 
\affiliation{Department of Astronomy, School of Physics, Peking University, Beijing 100871, China} 

\author{Y.~Homayouni}
\affiliation{Department of Physics, University of Connecticut, 2152 Hillside Rd Unit 3046, Storrs, CT 06269, USA}

\author{Jennifer~I-Hsiu~Li}
\affiliation{Department of Astronomy, University of Illinois at Urbana-Champaign, Urbana, IL 61801, USA}

\author{Ian~D.~McGreer}
\affiliation{Steward Observatory, The University of Arizona, 933 North Cherry Avenue, Tucson, AZ 85721, USA}

\author{B.~M.~Peterson}
\affiliation{Department of Astronomy, The Ohio State University, 140 W 18th Avenue, Columbus, OH 43210, USA}
\affiliation{Center for Cosmology and AstroParticle Physics, The Ohio State University, 191 West Woodruff Avenue, Columbus, OH 43210, USA}
\affiliation{Space Telescope Science Institute, 3700 San Martin Drive, Baltimore, MD 21218, USA }


\author{Dmitry Bizyaev}
\affiliation{Apache Point Observatory and New Mexico State University, P.O. Box 59, Sunspot, NM, 88349-0059, USA}
\affiliation{Sternberg Astronomical Institute, Moscow State University, Moscow, Russia}

\author{Yuguang Chen}
\affiliation{California Institute of Technology, 1200 E California Blvd., MC 249-17, Pasadena, CA 91125, USA}  

\author{K.~S.~Dawson}
\affiliation{Department of Physics and Astronomy, University of Utah, 115 S. 1400 E., Salt Lake City, UT 84112, USA} 

\author{Sarah~Eftekharzadeh}
\affiliation{Department of Physics and Astronomy, University of Utah, 115 S. 1400 E., Salt Lake City, UT 84112, USA}

\author{P.~J. Green}
\affiliation{Harvard-Smithsonian Center for Astrophysics, 60 Garden Street, Cambridge, MA 02138, USA} 

\author{Yucheng Guo}
\affiliation{Department of Astronomy, School of Physics, Peking University, Beijing 100871, China} 

\author{Siyao~Jia}
\affiliation{Department of Astronomy, University of California, Berkeley, CA 94720, USA}

\author{Linhua~Jiang}
\affiliation{Kavli Institute for Astronomy and Astrophysics, Peking University, Beijing 100871, China}

\author{Jean-Paul Kneib}
\affiliation{Institute of Physics, Laboratory of Astrophysics, Ecole Polytechnique F\'ed\'erale de Lausanne (EPFL), Observatoire de Sauverny, 1290 Versoix, Switzerland}
\affiliation{Aix Marseille Universit\'e, CNRS, LAM (Laboratoire d'Astrophysique de Marseille) UMR 7326, 13388, Marseille, France}

\author{Feng Li} 
\affiliation{School of Mathematics and Physics, Changzhou University, Changzhou 213164, China} 

\author{Zefeng Li} 
\affiliation{Department of Astronomy, School of Physics, Peking University, Beijing 100871, China} 

\author{Jundan Nie}
\affiliation{Key Laboratory of Optical Astronomy, National Astronomical Observatories, Chinese Academy of Sciences, Beijing 100012, China}

\author{Audrey Oravetz}
\affiliation{Apache Point Observatory and New Mexico State University, P.O. Box 59, Sunspot, NM, 88349-0059, USA}

\author{Daniel Oravetz}
\affiliation{Apache Point Observatory and New Mexico State University, P.O. Box 59, Sunspot, NM, 88349-0059, USA}

\author{Kaike Pan}
\affiliation{Apache Point Observatory and New Mexico State University, P.O. Box 59, Sunspot, NM, 88349-0059, USA}

\author{Patrick Petitjean} 
\affiliation{Institut d'Astrophysique de Paris, Sorbonne Universit\'e and CNRS,
   98bis Boulevard Arago, 75014, Paris, France}

\author{Kara~A.~Ponder}
\affiliation{Berkeley Center for Cosmological Physics,
    University of California Berkeley,
    341 Campbell Hall, Berkeley, CA 94720, USA  }

\author{Jesse~Rogerson}
\affiliation{Canada Aviation and Space Museum, 11 Aviation Parkway, Ottawa, ON, K1K 4Y5, Canada}
\affiliation{Department of Physics and Astronomy, York University, Toronto, ON M3J 1P3, Canada}

\author{M.~Vivek} 
\affiliation{Department of Astronomy and Astrophysics, Eberly College of Science, The Pennsylvania State University, 525 Davey Laboratory, University Park, PA 16802}
\affiliation{Institute for Gravitation \& the Cosmos, The Pennsylvania State University, University Park, PA 16802}

\author{Tianmen Zhang} 
\affiliation{Key Laboratory of Optical Astronomy, National Astronomical Observatories, Chinese Academy of Sciences, Beijing 100012, China}
\affiliation{School of Astronomy and Space Science, University of Chinese Academy of Sciences}

\author{Hu Zou} 
\affiliation{Key Laboratory of Optical Astronomy, National Astronomical Observatories, Chinese Academy of Sciences, Beijing 100012, China}

\begin{abstract}
We investigate the effects of extended multi-year light curves (9-year photometry and 5-year spectroscopy) on the detection of time lags between the continuum variability and broad-line response of quasars at $z\gtrsim 1.5$, and compare with the results using 4-year photometry$+$spectroscopy presented in a companion paper. We demonstrate the benefits of the extended light curves in three cases: (1) lags that are too long to be detected by the shorter-duration data but can be detected with the extended data; (2) lags that are recovered by the extended light curves but missed in the shorter-duration data due to insufficient light curve quality; and (3) lags for different broad line species in the same object. These examples demonstrate the importance of long-term monitoring for reverberation mapping to detect lags for luminous quasars at high-redshift, and the expected performance of the final dataset from the Sloan Digital Sky Survey Reverberation Mapping project that will have 11-year photometric and 7-year spectroscopic baselines. 

\end{abstract}


\section{Introduction} \label{sec:introduction}

Reverberation mapping \citep[RM, e.g.,][]{Blandford_McKee_1982,Peterson_2014} is a technique to measure the time delay between the continuum variability and the response of the broad emission lines powered by the continuum in efficiently accreting supermassive black holes (referred to as ``quasars'' throughout this paper). This time delay reflects the light travel time from the innermost region around the black hole to the broad-line region (BLR) and hence measures a characteristic size for the BLR. Combined with the virial velocity of the BLR inferred from the width of the broad emission lines, RM is used as the primary method to estimate the black hole masses of distant quasars \citep[][]{Peterson_2014}. 

Traditionally, RM programs require dedicated imaging and spectroscopic resources, usually obtained from small telescopes with guaranteed access, to monitor individual objects over an extended period of time (typically several months) where the variability and the lag can be temporally resolved. Because of the required heavy commitment of monitoring resources and the serial mode of observing, past RM work has been mostly limited to the low-$z$ and low-luminosity regime of quasars, targeting the most variable nearby systems for which RM is most likely to succeed \citep[e.g.,][]{Kaspi_etal_2000,Peterson_etal_2004,Barth_etal_2015,Du_etal_2016}, with a few exceptions for high-redshift and high-luminosity quasars \citep[e.g.,][]{Kaspi_etal_2007,Trevese_etal_2014,Lira_etal_2018,Czerny_etal_2019}. To expand the RM sample to a broader parameter space of quasars including the high-$z$ and high-luminosity regime, the most efficient approach is to monitor a significant number of quasars simultaneously with wide-field imaging and multi-object spectroscopy. 

The Sloan Digital Sky Survey Reverberation Mapping (SDSS-RM) project is a dedicated multi-object optical reverberation mapping program \citep{Shen_etal_2015a} that has been monitoring a single 7 ${\rm deg}^2$ field since 2014, using the SDSS Baryon Oscillation Spectroscopic Survey \citep[BOSS,][]{Eisenstein_etal_2011,Dawson_etal_2013} spectrographs  \citep[][]{Smee_etal_2013} on the 2.5m SDSS telescope \citep{Gunn_etal_2006} at Apache Point Observatory. Accompanying photometric data in the $g$ and $i$ bands are acquired with the 3.6 m Canada-France-Hawaii Telescope (CFHT) and the Steward Observatory 2.3 m Bok telescope. By the end of SDSS-RM in 2020, the spectroscopic baseline will span 7 years, with cadences ranging from 4-5 days in the first season to monthly in the last few seasons (2018-2020); the photometric baseline will span 11 years when including the 2010-2013 photometric light curves from the Pan-STARRS 1 \citep[PS1,][]{Kaiser_etal_2010} Medium Deep survey that covers the entire SDSS-RM field. The SDSS-RM sample includes 849 quasars with $i_{\rm SDSS}<21.7$ and $0.1<z<4.5$ without any constraints on quasar properties \citep[for detailed sample properties, see][]{Shen_etal_2019b}. The main purpose of SDSS-RM is to measure RM lags of different broad lines covered by optical spectroscopy across the full range of quasar luminosity and redshift probed by the sample. 

SDSS-RM has successfully measured short ($<6\, {\rm months}$) lags for the low-ionization broad lines \citep[e.g., \halpha, \hbeta, and \MgII,][]{Shen_etal_2016a,Grier_etal_2017} based on the 2014 data. In a companion paper \citep[][hereafter Paper I]{Grier_etal_2019}, we reported the first results on \CIV\ lags using four years (2014-2018) of imaging and spectroscopy from SDSS-RM (and for \MgII\ lags in Homayouni, Y., et~al., in prep), where the lags are typically longer than one season in the observed frame. In this work, we investigate the benefits of extending the baseline for the detection of lags. Because the calibration and preparation of multi-year light curves for lag measurements is computationally demanding (see \S\ref{sec:data}), we highlight only a handful of selected objects to demonstrate the power of extended baselines. A complete investigation of long lags for the full SDSS-RM sample will be reported in future work. We describe the extended light curve data used in \S\ref{sec:data}, present the lag measurements in \S\ref{sec:analysis}, and discuss our results and conclude in \S\ref{sec:con}. Luminosities are calculated assuming a flat $\Lambda$CDM cosmology with parameters $\Omega_{\Lambda}=0.7$, $\Omega_{\rm m}=0.3$, and $H_0=70$ km s$^{-1}$ Mpc$^{-1}$.

\section{Data}\label{sec:data}

\begin{figure*}
 \includegraphics[width=0.46\textwidth]{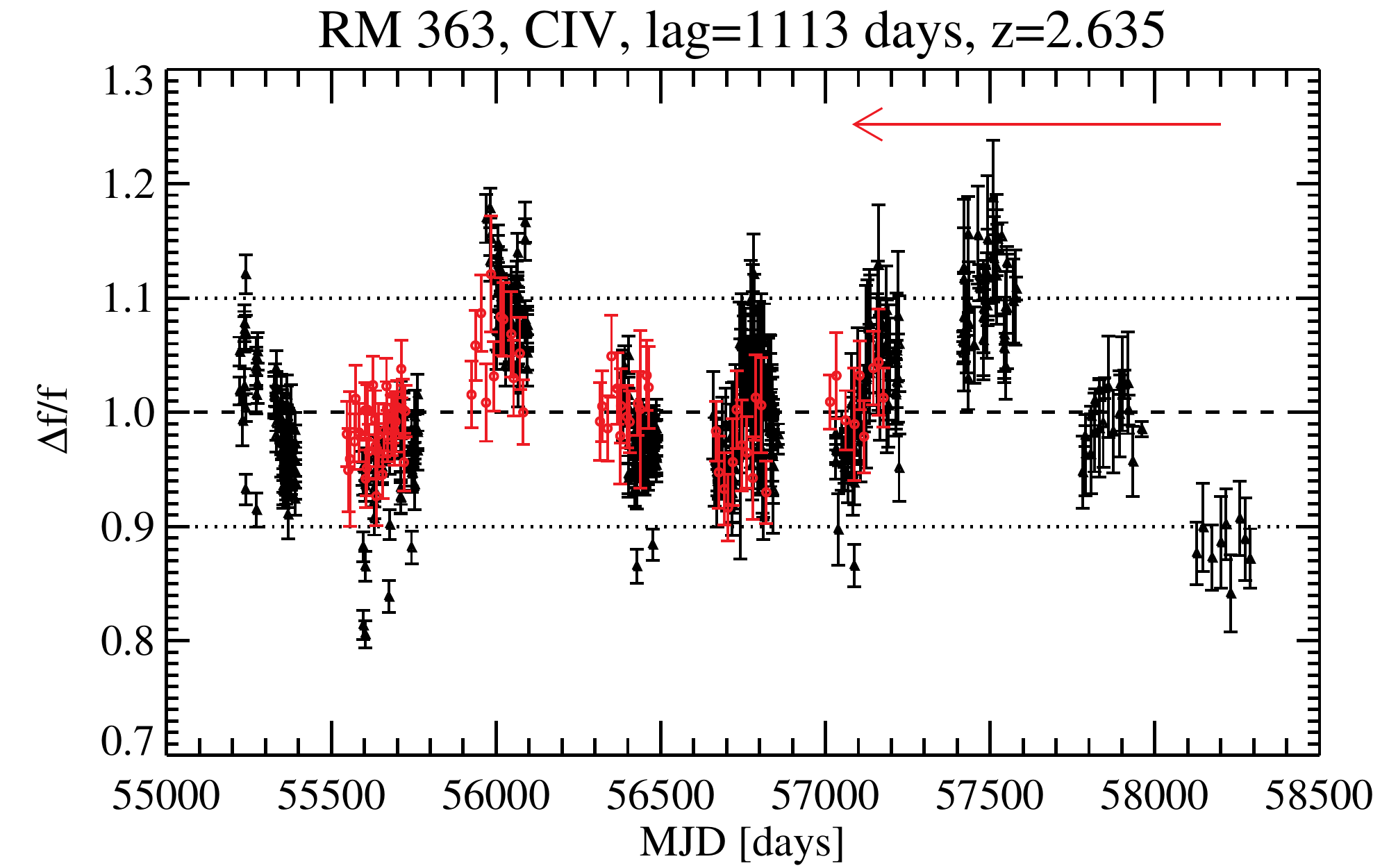}
 \includegraphics[width=0.46\textwidth]{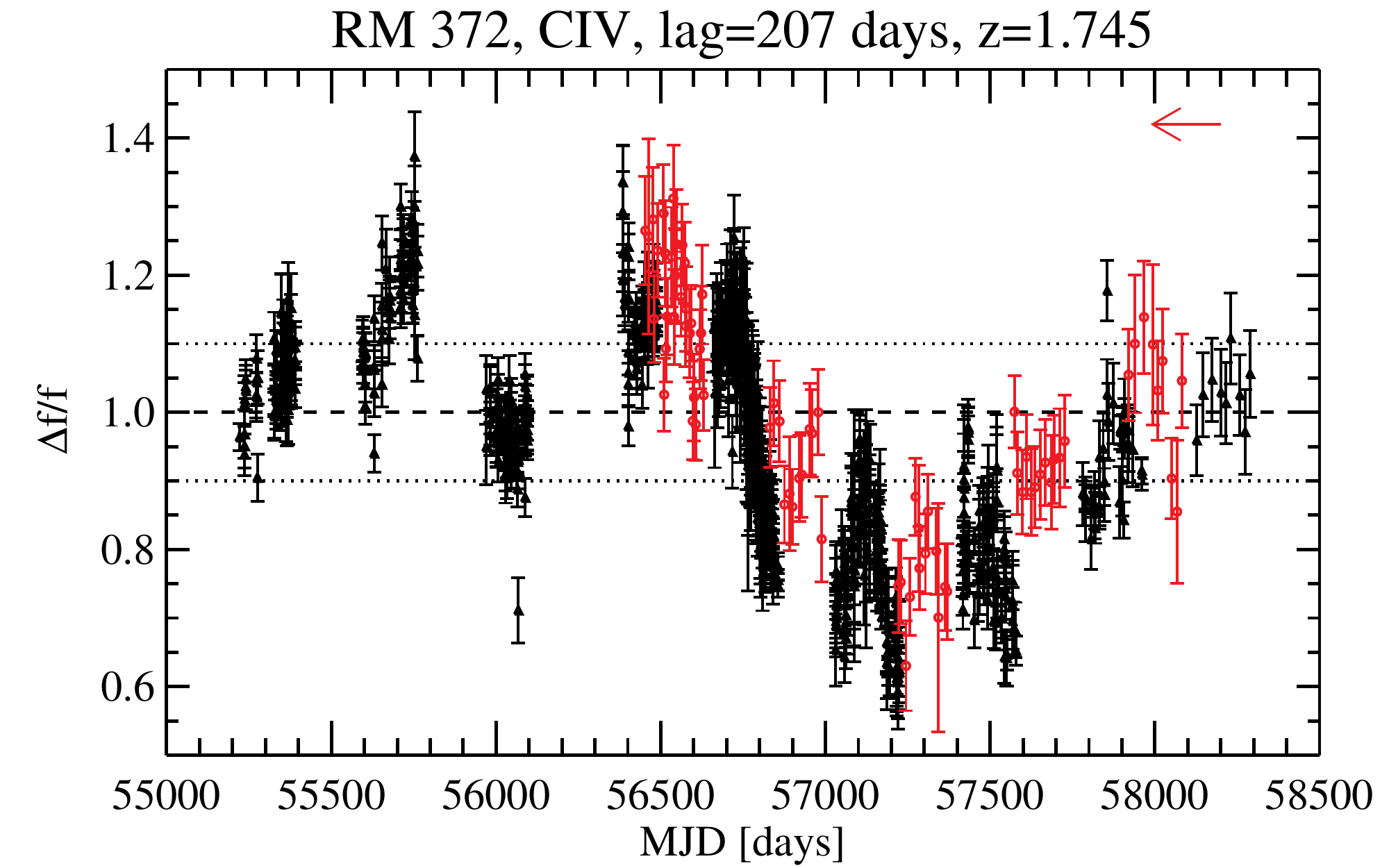}\\
 \includegraphics[width=0.46\textwidth]{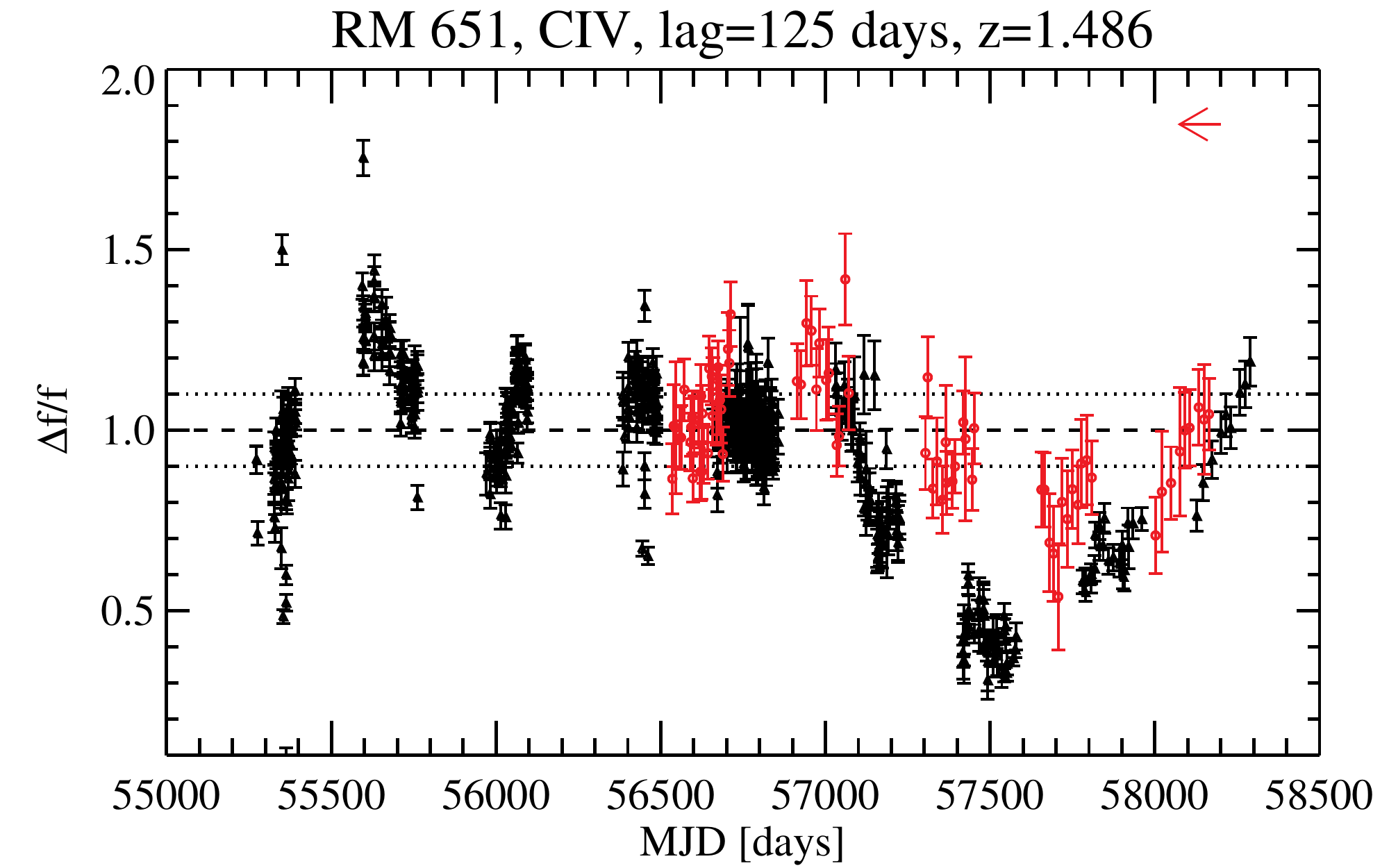}
 \includegraphics[width=0.46\textwidth]{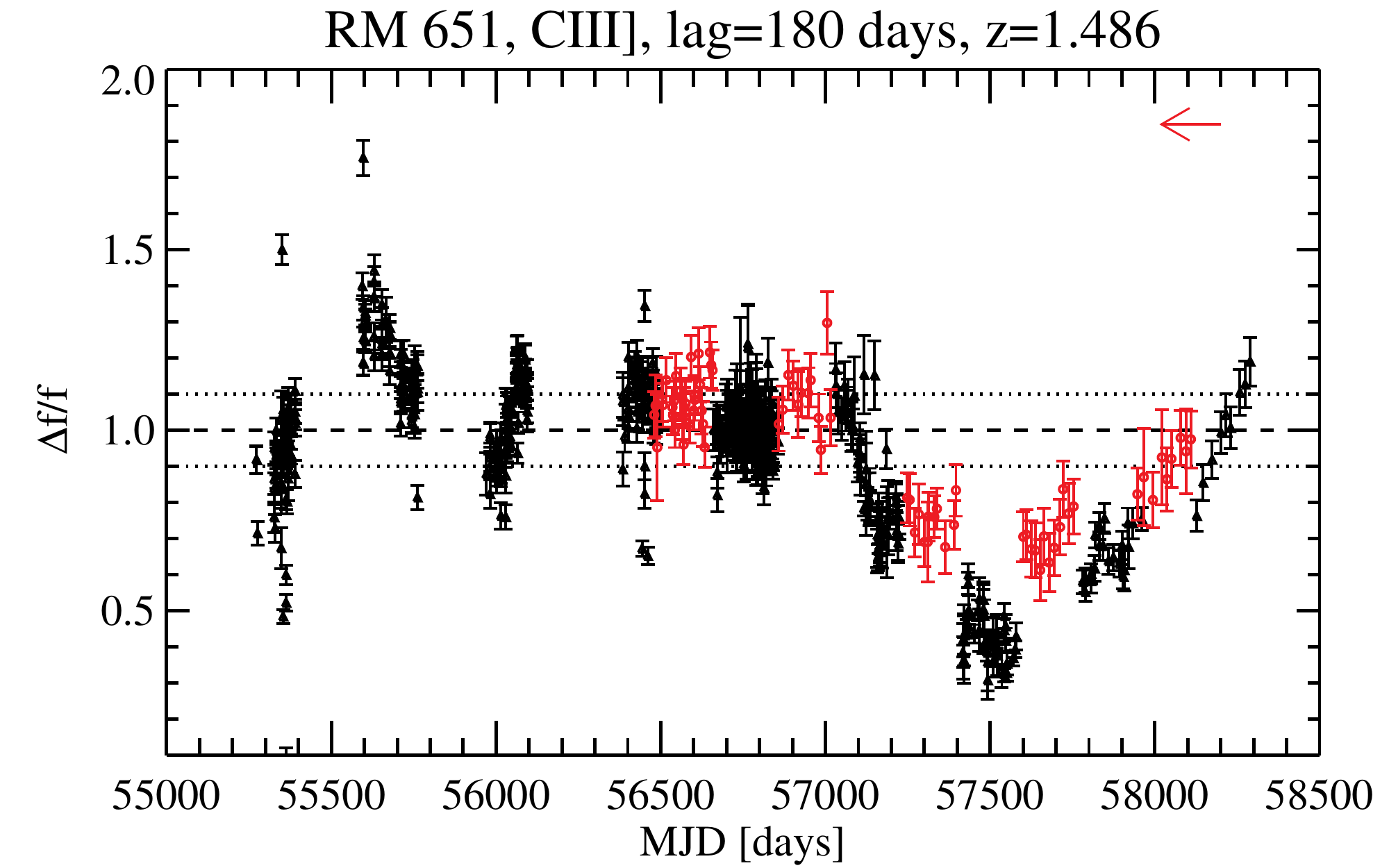}\\
 \includegraphics[width=0.46\textwidth]{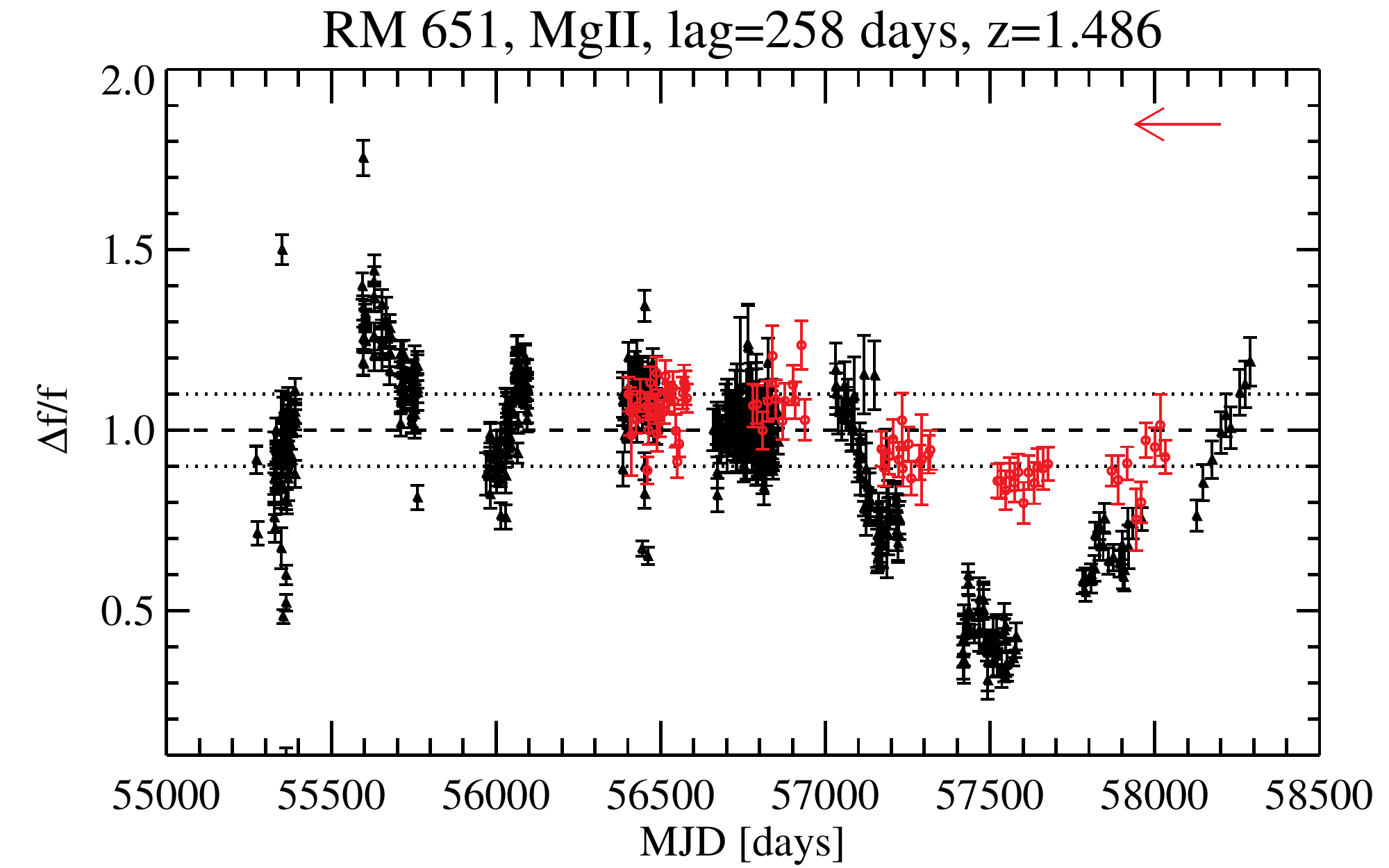}
 \caption{The 9-year $g$-band continuum light curve (black) and 5-year line light curve (red) for different objects and lines, where the latter has been shifted by the corresponding {\tt JAVELIN} lag (see summary in Table \ref{tab:summary}), indicated by the red arrow in each panel. The light curves are normalized using the median flux over the period to reflect fractional changes. All available epochs (including multiple data points on the same night) are plotted. } \label{fig:lc}
 \end{figure*}

\begin{figure*}
 \includegraphics[width=0.465\textwidth]{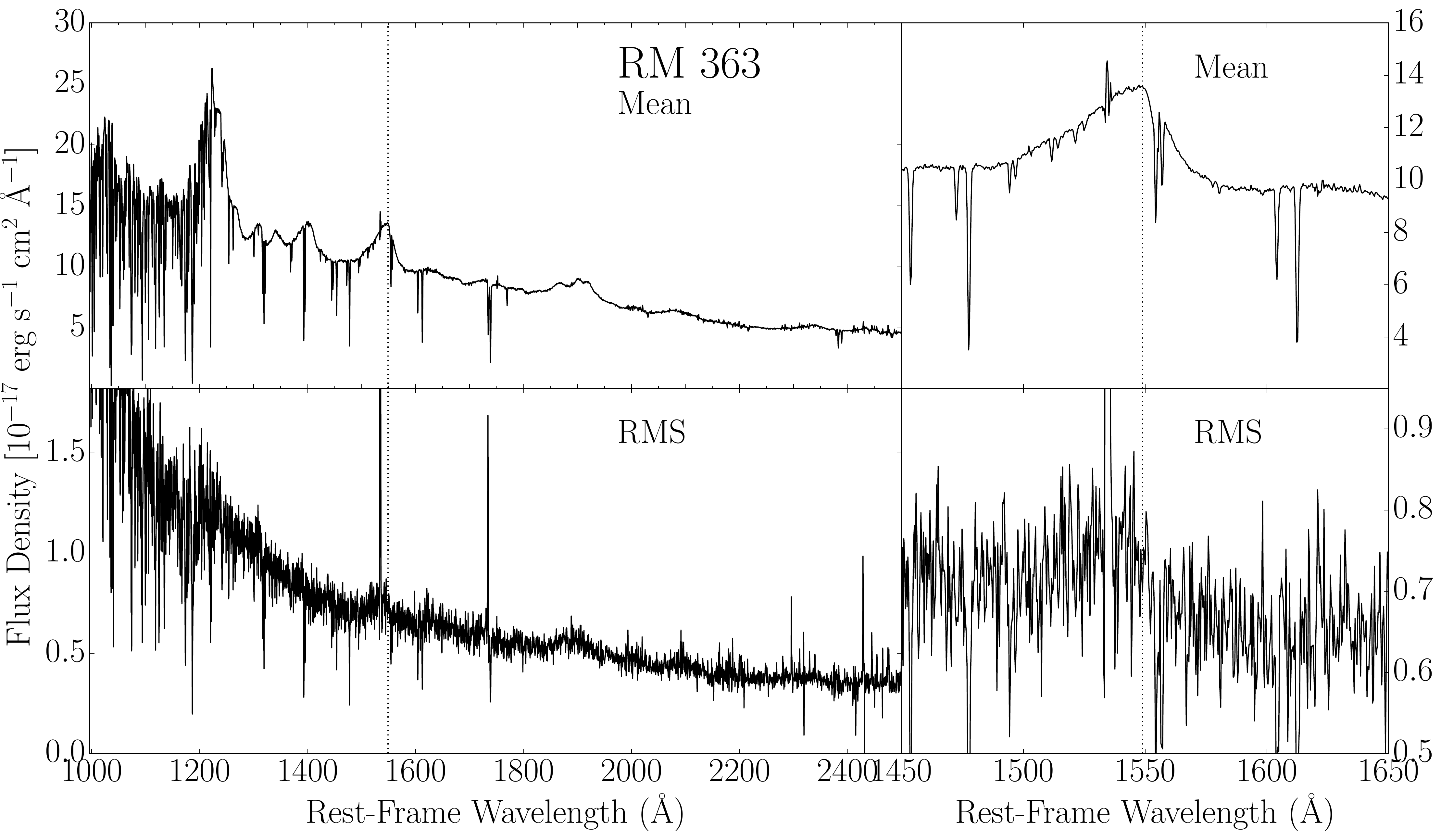}\hspace{0.5cm}
  \includegraphics[width=0.465\textwidth]{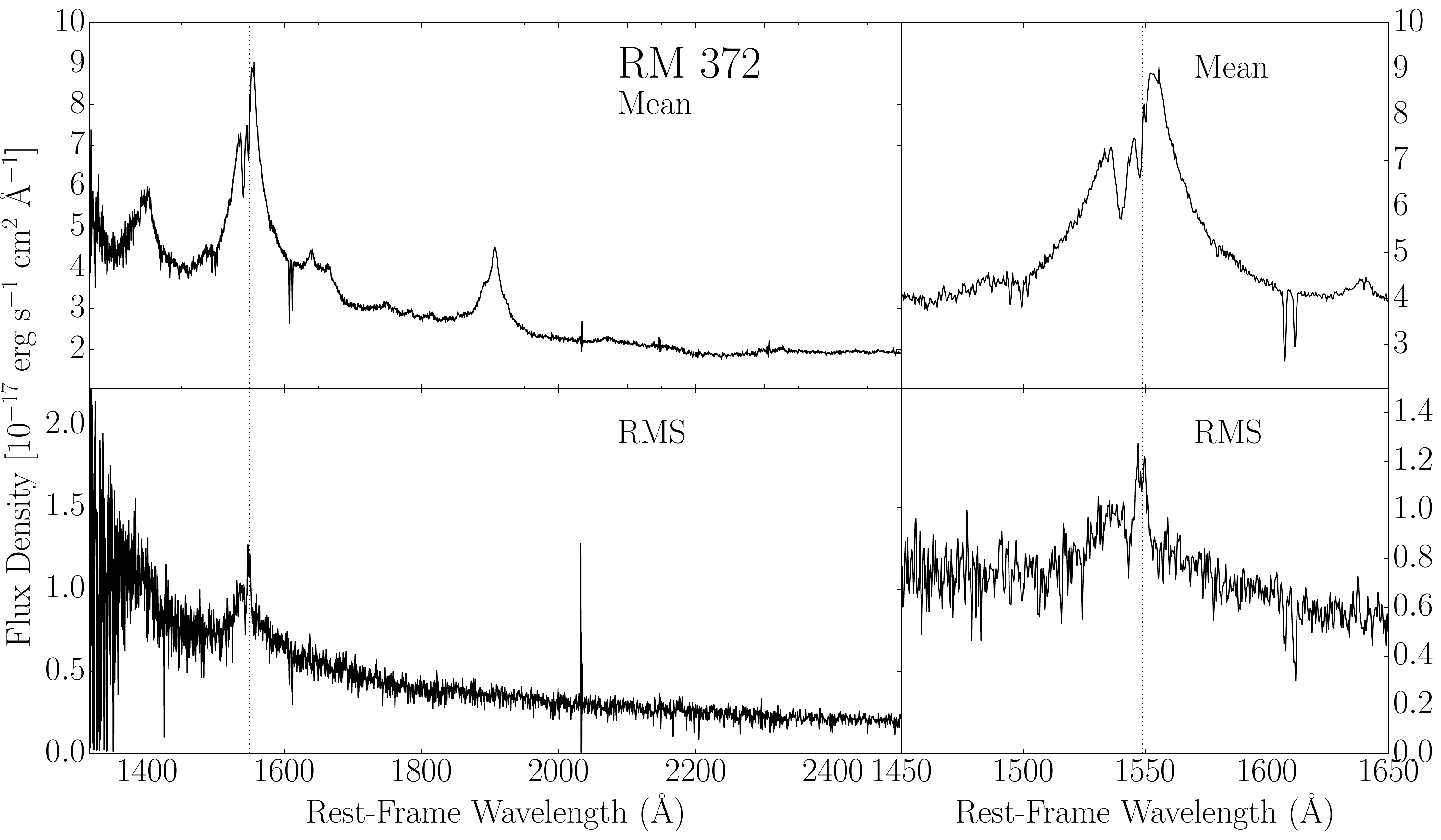}\\
   \includegraphics[width=0.95\textwidth]{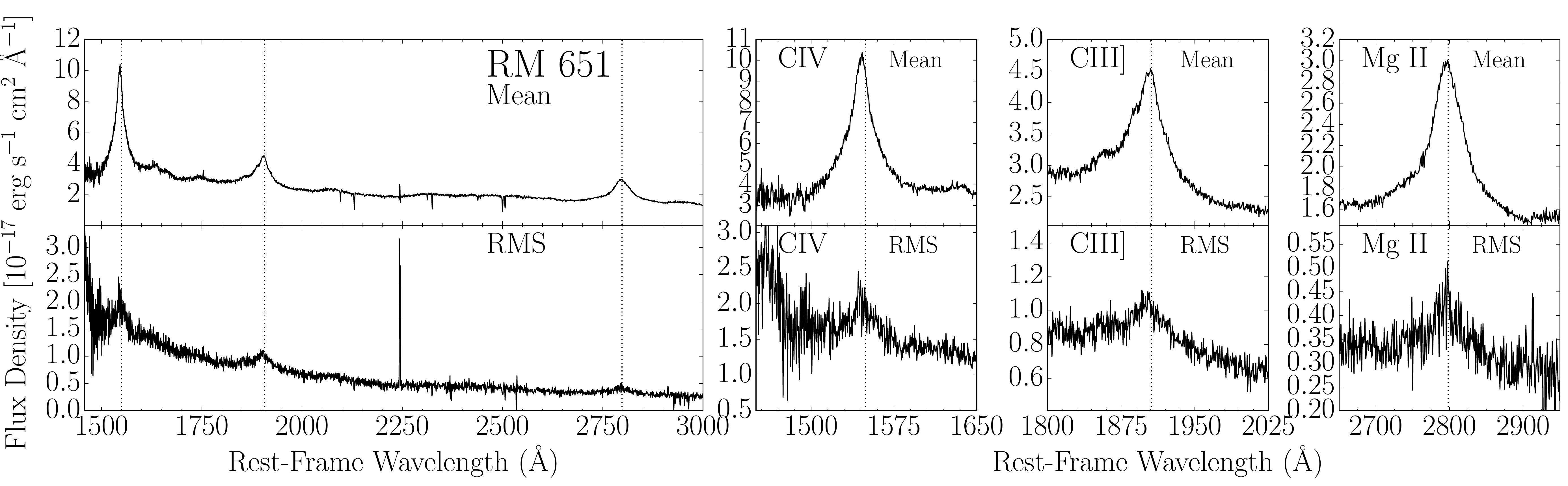}
 \caption{Mean and RMS spectra for the three objects in this work. The top row is for RM 363 and RM 372 (for \CIV\ lag), and the bottom row is for RM 651 (for three lines). For each object, the top and bottom panels show the mean and RMS spectra. The left panels show a large portion of the observed spectrum, and the right panels show only the emission-line region. Vertical dotted black lines indicate the rest-frame wavelength of the broad emission line.} \label{fig:rms}
 \end{figure*}

The SDSS-RM monitoring data include the multi-epoch spectroscopy taken by the BOSS spectrographs, as well as photometric light curves from the CFHT and Bok telescopes. The spectroscopic data are first re-processed with a custom pipeline to improve flux calibration \citep[][]{Shen_etal_2015a}, followed by another recalibration process called PrepSpec to further enhance spectrophotometric accuracy using the flux of the narrow emission lines \citep[e.g.,][]{Shen_etal_2016a,Grier_etal_2017}. PrepSpec also produces the broad-line light curves and the RMS spectra computed from the multi-epoch spectroscopy, which are used in subsequent analysis. 

The photometric data from Bok and CFHT are combined with the synthetic $g$ and $i$ photometry computed from SDSS spectroscopy. The light curve merging is performed using the Continuum REprocessing AGN MCMC ({\tt CREAM}) software developed by \citet{Starkey_etal_2016} to produce the continuum light curves for the lag measurements. During the {\tt CREAM} merging process, all datasets in both ($g$ and $i$) filters are scaled to a common flux scale of the $g$-band synthetic photometry computed from SDSS spectra, and the uncertainties of the flux measurements for both the continuum and broad-line light curves are adjusted automatically to account for possible underestimation of the flux errors. {\tt CREAM} typically increases the flux uncertainties by a fraction of less than $50\%$, indicating that the original uncertainties are usually reasonable, but we prefer the more conservative flux uncertainties reported by {\tt CREAM}. More technical details of the {\tt CREAM} procedure can be found in Paper I. The PS1 data are not combined using {\tt CREAM} with the rest of the light curve because of the lack of overlap in time coverage, and are simply converted to flux density at the effective wavelength of $g$ band. There is no noticeable offset in the $g$-band light curve between the PS1 and CREAM-merged datasets based on standard stars. We use the combined 9-yr $g$-band data as our fiducial continuum light curve to measure the light curve statistics and broad-line lags. 
 
 \begin{figure*}
 \includegraphics[width=0.95\textwidth]{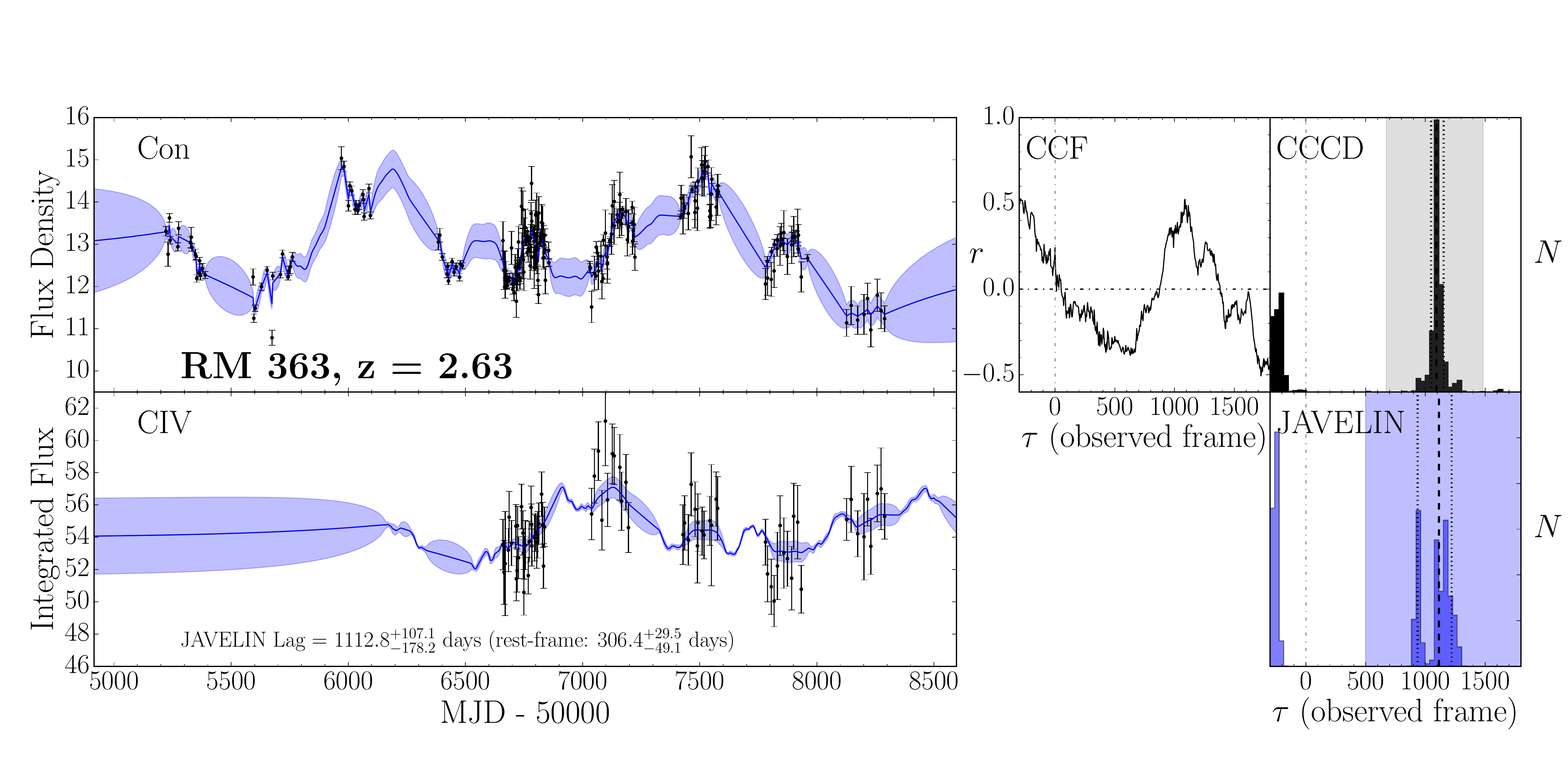}\vspace{-1cm}
 \includegraphics[width=0.95\textwidth]{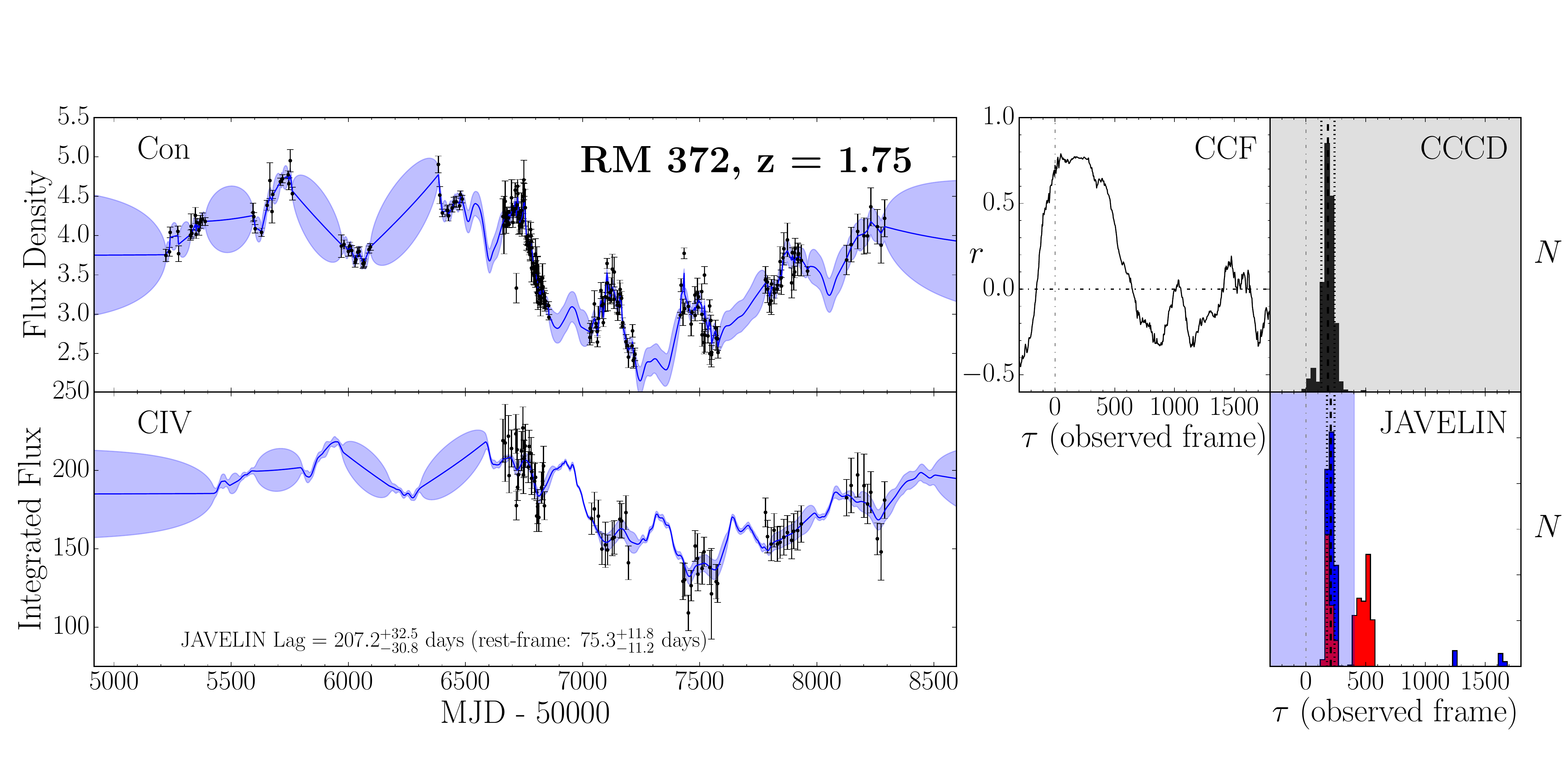}
 \caption{Lag measurements based on the 9-year continuum light curve and 5-year line light curve for RM 363 and RM 372. For each object, the left two panels show the nightly-averaged light curves for the continuum and the emission line, respectively. The nightly-binned data are for plotting purposes only, and the lag measurements are performed with all data points. The blue line and the shaded area are the best-fit {\tt JAVELIN} continuum model light curve and $1\sigma$ uncertainties. The right three panels present the ICCF, the distribution of the CCF centroid (CCCD) from Monte Carlo resampling of the light curves, and the posterior distribution of lags from the {\tt JAVELIN} analysis. In the {\tt JAVELIN} panel (lower-right), the light blue-shaded area is the region used to compute the lag and uncertainties, following the alias-mitigation approach described in detail in Paper I \citep[also see][]{Grier_etal_2017}. The dashed and dotted lines indicate the lag (measured from the median of the distribution within the shaded area) and its $1\sigma$ uncertainties. For RM 363, The inclusion of the earlier photometry from PS1 prior to 2014 and the fifth year of the spectroscopic light curve are critical to this measurement. For RM 372,  we were able to measure a robust \CIV\ lag, while the {\tt JAVELIN} analysis in Paper I based on 4 years of photometry and spectroscopy (shown in red in the {\tt JAVELIN} panel) has an alias peak around 500 days and a lower significance for the correct lag.  \label{fig:rm363}}
 \end{figure*}


Paper I analyzed the 4-year data set for a sample of 349 quasars with \CIV\ coverage ($1.35<z<4.32$) and significant line variability detected by PrepSpec. We are in the process of merging light curves with the latest compilation of data (including additional years of spectroscopy and photometry), and the {\tt CREAM} step is the most computationally demanding step in the process. We thus focus this study on a handful of cases of particular interest for which we have prioritized the light curve processing, using 5 years of spectroscopy (2014-2018) and 9 years of photometry (2010-2018). The leading photometric light curves from PS1 that do not have accompanying spectroscopy effectively extend the baseline to 9 years. This work uses the PS1 data presented by \citet{Shen_etal_2019b} for the lag analysis. 

\section{Lag Measurements}\label{sec:analysis}

With the {\tt CREAM}-merged light curves, we follow the procedure outlined in Paper I to measure the lag between the continuum and the broad-line light curves. We have performed a systematic comparison (Li \etal\ 2019) of the performance of several commonly-adopted methods to measure the lag, including the Interpolated Cross-Correlation Function \citep[ICCF,][]{Gaskell_Peterson_1987}, $z$-Transformed Discrete Correlation Function \citep[ZDCF,][]{Alexander_2013}, and {\tt JAVELIN} \citep{Zu_etal_2011}. For data with quality similar to SDSS-RM data, {\tt JAVELIN} provides the best performance in terms of lag recovery and the fidelity of the lag uncertainties. We therefore adopt the lags measured by {\tt JAVELIN} as our fiducial lags. However, ICCF has been the standard technique to measure RM lags in the literature, so we also present results from the ICCF approach to validate the reported lags. 

We are particularly interested in cases where the extended monitoring data improve the lag detection over the 4-year data used in Paper I. More data points and extended baseline enhances the correlation between the light curves, mitigates aliases from insufficient sampling, and allows us to measure lags that are too long to be detectable by the shorter-duration RM data. We here present results for three sources selected from the 349 quasars in Paper I to demonstrate the power of the extended light curves. These were chosen from a handful of objects with well behaved 4-yr light curves for which we compiled the extended light curves and search for lags. We imposed no additional criteria on these measurements for selection --- for example, we did not require that the lags are consistent with the expectation from the preliminary \CIV\ $R-L$ relation in Paper~I.   

The continuum and emission-line light curves for these examples are shown in Fig.\ \ref{fig:lc}; we discuss these individual cases in the following subsections. In all cases, the variability of the lines is well detected, and we show the mean and RMS spectra of these examples in Fig.~\ref{fig:rms}.
 
  
\begin{figure*}
  \includegraphics[width=0.95\textwidth]{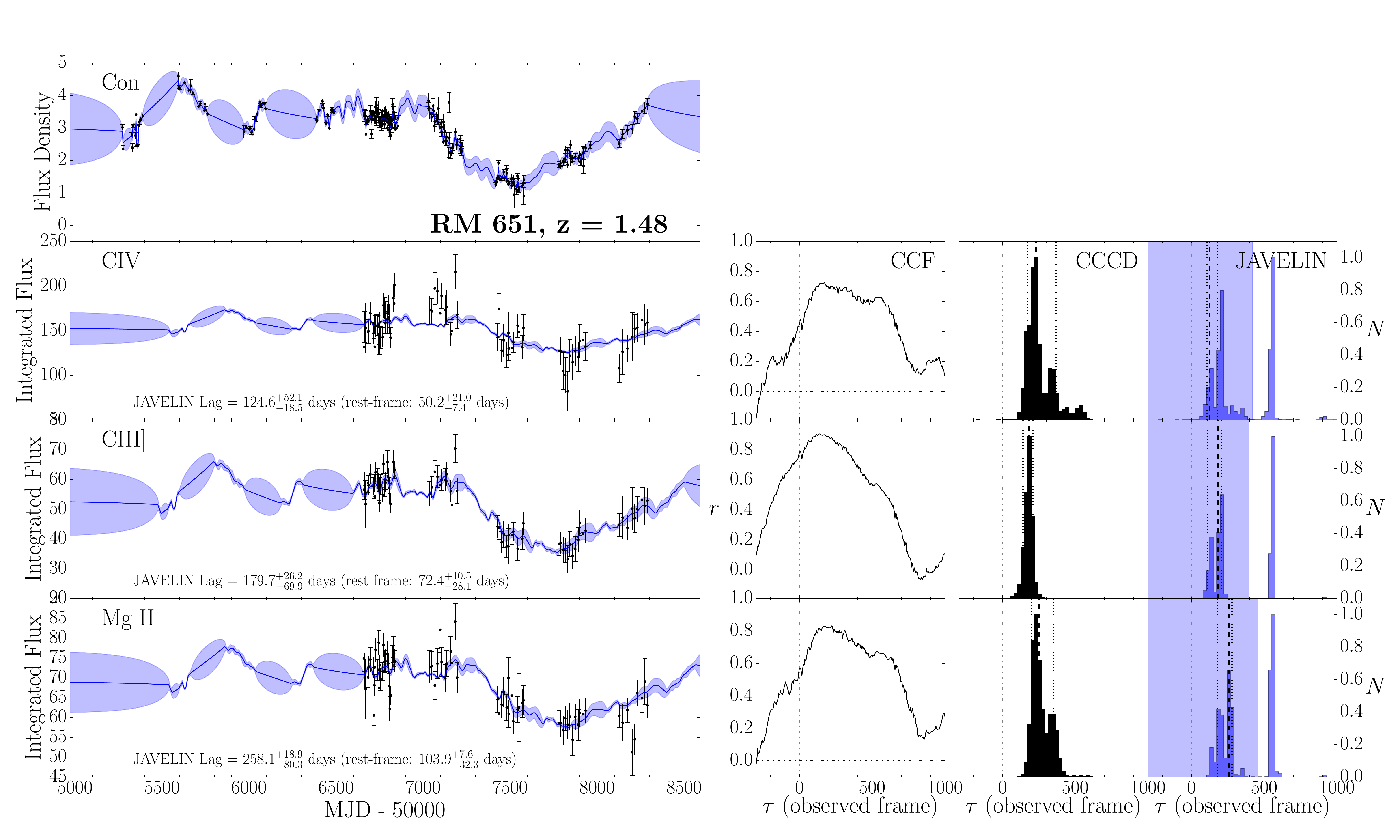}
  \caption{Same as Fig.\ \ref{fig:rm363} but for RM 651 and for three broad lines, \CIV, \CIII\ and \MgII. We successfully recovered lags in all three broad lines. The \CIII\ line flux includes contributions from the \SiIII\ and \AlIII\ lines. \label{fig:rm651}}
 \end{figure*}  
  
\subsection{RM 363: detecting longer \CIV\ lags}

The range of lag sensitivity of a campaign is limited by the baseline of the campaign; the delayed broad-line light curve must still have substantial overlap with the leading continuum light curve in order to cross-correlate the two light curves. The data examined in Paper I span only 4 years; Paper I was thus limited to lags shorter than about 700 days in the observed frame.

Here, with the 9-year effective baseline, we extend our lag search to 1800 days (roughly $60\%$ of the baseline), suitable for detecting longer lags in the more luminous and higher redshift subset of the SDSS-RM sample. 

Fig.\ \ref{fig:rm363} (top) presents such an example: we measure a \CIV\ lag in RM 363, one of the most luminous quasars in the SDSS-RM sample, having a bolometric luminosity of $\sim 10^{47}\,{\rm erg\,s^{-1}}$ at $z=2.635$. The measured \CIV\ lag is more than 1000 days in the observed frame due to the $1+z$ time dilation. This example demonstrates the necessity of long-term monitoring for the detection of multi-year lags. In addition, the typical variability amplitude of this high-redshift quasar is only $\sim 10\%$ (see Fig.\ \ref{fig:lc}), indicating that high S/N spectroscopy is required to detect the line variability. 

\subsection{RM 372: mitigating for aliases}

Fig.\ \ref{fig:rm363} (bottom) presents our RM analysis for RM 372 at $z=1.745$. This source was analyzed in Paper I with the 4-year light curves, and a \CIV\ lag was not robustly detected due to insufficient light curve quality. With the addition of the 5$^{\rm th}$ year of spectroscopy and the 4 years of PS1 monitoring, we were able to recover an observed-frame lag of $\sim 200$ days. 

The \CIV\ line light curve is shifted by about 6 months from the continuum light curve, resulting in little overlap in the two light curves. The light curves have smooth variations over multi-year timescales, and the correlations between the continuum and line light curves are strongly suppressed on timescales longer and shorter than $\sim 6$ months, allowing a time lag of about the length of the seasonal gap to be robustly detected. This example demonstrates the power of extended-baseline light curves in distinguishing between alias lags and true lags, and shows that when certain conditions are met, we are able to measure lags that are usually difficult to measure when there are regular seasonal gaps. However, if the variations dominating the correlation in the light curves are on timescales shorter than $\sim 1$ yr, the seasonal gaps will likely lead to unreliable lag detection or even no detection \citep[e.g.,][]{Grier_etal_2008}. 


\subsection{RM 651: lags from multiple lines}

Given the broad spectral coverage of SDSS, we will be able to measure lags from several broad lines in a single object, and explore the stratification of the broad-line region \citep[e.g.,][]{Peterson_Wandel_1999}. Fig.\ \ref{fig:rm651} presents the analysis for RM 651 at $z=1.486$, where lags are successfully detected for three different lines: \CIV, \CIII\ and \MgII. For \CIII, PrepSpec measures the line flux from the complex of \CIII, \AlIII\ and \SiIII, where \CIII\ dominates the flux. For this particular object, there is some evidence from the {\tt JAVELIN} analysis that the \CIV\ lag $<$ \CIII\ lag $<$ \MgII\ lag, which is suggestive of BLR stratification. However, the lag uncertainties are large, and the line widths measured from the rms spectrum do not differ significantly for these three lines (see Table \ref{tab:summary}). Therefore we cannot confirm BLR stratification for this example. 

The \MgII\ variability amplitude in this object is substantially lower than that of the continuum and the other two lines (see Fig.~\ref{fig:lc}). This behavior is generally consistent with the lower variability of \MgII\ compared to other major broad emission lines \citep[e.g.,][]{Sun_etal_2015}, and also expectations from photoionization models \citep[e.g.,][]{Guo_etal_2019}.

\section{Discussion and Conclusions}\label{sec:con}

Using a 9-year photometric baseline and 5-year spectroscopic baseline, we have presented several examples of lag detections in high-redshift ($z\gtrsim 1.5$) quasars from the SDSS-RM project. We highlighted three objects: (1) a luminous quasar with an observed-frame \CIV\ lag of more than 1000 days; (2) a quasar with a lag that falls within the seasonal gap of the light curves that was recovered by the extended light curves but missed in the earlier analysis based on 4-year light curves (Paper I); and (3) a quasar with well-detected lags for three broad lines (\CIV, \CIII\ and \MgII). 

\begin{figure}
  \includegraphics[width=0.48\textwidth]{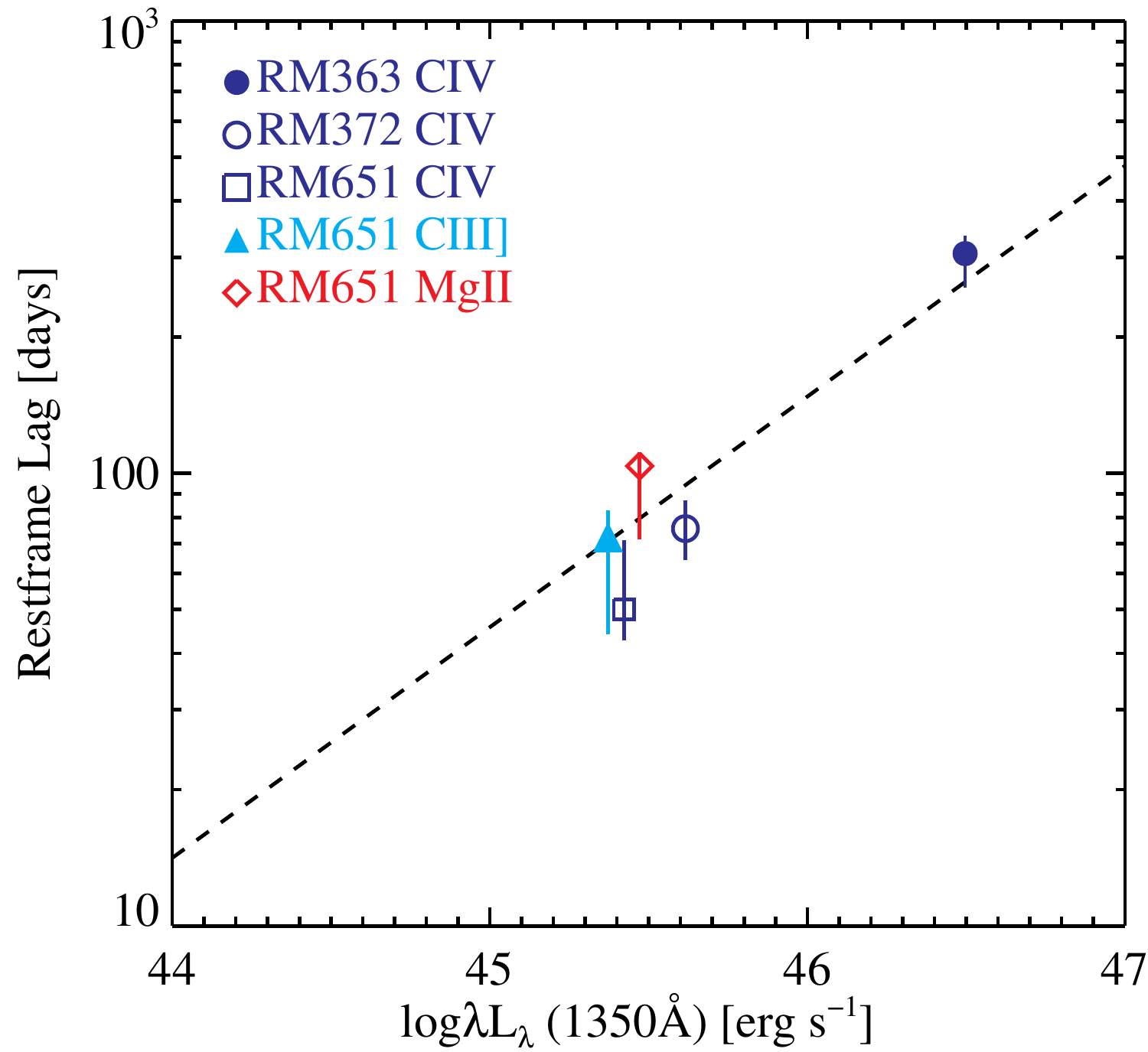}
  \caption{Locations of the measured rest-frame lags for our example cases in the $R-L$ plane. For RM 651, the luminosity positions have been slightly shifted for different lines for clarity. The dashed line is the best-fit result in Paper I for \CIV\ lags. The measured lags are all consistent with this relation. There is a hint that the \CIV\ (\MgII) lag is the shortest (longest) among the three lines, but the uncertainties are too large to confirm this trend.  \label{fig:RL}}
 \end{figure}

Fig.~\ref{fig:RL} displays the locations of our measured lags in the radius-luminosity ($R-L$) plane. These lags are consistent with the best-fit $R-L$ relation for \CIV\ based on the full sample in Paper I. However, we caution that the $R-L$ relation for \CIV\ is still poorly constrained at the moment, a situation that can only be improved by obtaining more \CIV\ lag measurements over a large dynamic range in quasar luminosity \citep[e.g.,][Paper I]{Lira_etal_2018,Hoormann_etal_2019}. 

This work demonstrates the benefits, and sometimes the necessity, of having long baselines of multi-season RM data to detect long lags robustly in high-redshift and high luminosity quasars. Our preliminary analysis for a small number of objects provides a glimpse of the expected lags from the final data set of SDSS-RM, which will include two more years of photometric and spectroscopic monitoring. With the final extended dataset, we expect to detect lags in the most luminous quasars in the SDSS-RM sample, thus further expanding the dynamic range in luminosity to constrain the $R-L$ relation for \CIV. These results also demonstrate that lag measurements for multiple lines in the same object are possible with SDSS-RM data for high-redshift quasars, and hint at their possible utility in evaluating BLR stratification models.  

Our results here (and in Paper I) provide strong endorsement of current and future multiplexing spectroscopic RM programs, such as the Black Hole Mapper program in SDSS-V \citep{Kollmeier_etal_2017}, the OzDES RM program \citep{King_etal_2015}, the 4MOST RM program \citep{Swann_etal_2019}, and the Maunakea Spectroscopic Explorer RM program \citep{MSE2019}.  

\begin{table*}  
\caption {Summary of RM measurements} \label{tab:summary}
\begin{center}
\begin{tabular}{cc c c c c c c c} 
\hline \hline
RMID  & SDSS name & $z$ & $\log\lambda L_{\lambda} (1350\,{\textrm \AA})$ & line & $\tau_{\rm {\tt JAVELIN}}$ & $\tau_{\rm CCCD}$ & $\sigma_{\rm rms}$ & VP \\ 
      & hhmmss.ss$\pm$ddmmss.s &  & [${\rm erg\,s^{-1}}$]  & & (days) & (days) & (${\rm km\,s^{-1}}$) & ($10^8\,M_\odot$) \\
\hline
363 & 142113.30$+$524929.9 & 2.635  & 46.50 & \CIV & $1113_{-178}^{+107}$  &  $1092^{+62}_{-44}$ & $2230\pm 40$  & $3.0^{+0.3}_{-0.5}$ \\
372 &  141236.48$+$540152.1 & 1.745  & 45.62 & \CIV & $207_{-31}^{+33}$  & $184^{+56}_{-56}$  &  $4438\pm 21$ & $2.9^{+0.5}_{-0.4}$  \\
651 & 142149.30$+$521427.8 & 1.486 & 45.42$^a$ & \CIV & $125_{-19}^{+52}$  & $228^{+140}_{-58}$ & $2714\pm22$  & $0.72^{+0.30}_{-0.11}$ \\
       &  &         & & \CIII & $180_{-70}^{+26}$  & $180^{+30}_{-38}$ & $2645\pm 25 $ & $0.99^{+0.14}_{-0.39}$  \\
       &   &        & & \MgII & $258_{-80}^{+19}$  & $249^{+103}_{-50}$  & $2586\pm 29$ & $1.35^{+0.10}_{-0.42}$ \\
\hline
\end{tabular}
\tablecomments{The lags are reported in the observed frame. The last column lists the virial product ${\rm VP}\equiv \sigma_{\rm rms}^2c\tau_{\rm JAVELIN}/[G(1+z)]$, where $\sigma_{\rm rms}$ is the line dispersion measured from the line-only RMS spectrum based on the 5-year spectroscopy. $^a$The rest-frame 1350\,\AA\ is not covered in the spectrum of RM 651, and we compute $\log\lambda L_{\lambda} (1350\,{\textrm \AA})$ from the 1700\,\AA\ monochromatic luminosity assuming a typical power-law continuum of $f_\lambda\propto \lambda^{-1.5}$. }
\end{center}
\end{table*}

\acknowledgments

YS acknowledges support from an Alfred P. Sloan Research Fellowship and NSF grant AST-1715579. CJG, WNB, JRT, and DPS acknowledge support from NSF grants AST-1517113 and AST-1516784. KH acknowledges support from STFC grant ST/R000824/1. PBH acknowledges support from NSERC grant 2017-05983. LCH acknowledges National Science Foundation of China (11721303) and the National Key R\&D Program of China (2016YFA0400702).

This work is based on observations obtained with MegaPrime/MegaCam, a joint project of CFHT and CEA/DAPNIA, at the Canada-France-Hawaii Telescope (CFHT) which is operated by the National Research Council (NRC) of Canada, the Institut National des Sciences de l'Univers of the Centre National de la Recherche Scientifique of France, and the University of Hawaii.
The authors recognize the cultural importance of the summit of Maunakea to a broad cross section of the Native Hawaiian community. The astronomical community is most fortunate to have the opportunity to conduct observations from this mountain.

Funding for the Sloan Digital Sky Survey IV has been provided by the Alfred P. Sloan Foundation, the U.S. Department of Energy Office of Science, and the Participating Institutions. SDSS-IV acknowledges support and resources from the Center for High-Performance Computing at the University of Utah. The SDSS web site is www.sdss.org. SDSS-IV is managed by the Astrophysical Research Consortium for the Participating Institutions of the SDSS Collaboration including the Brazilian Participation Group, the Carnegie Institution for Science, Carnegie Mellon University, the Chilean Participation Group, the French Participation Group, Harvard-Smithsonian Center for Astrophysics, Instituto de Astrof\'isica de Canarias, The Johns Hopkins University, Kavli Institute for the Physics and Mathematics of the Universe (IPMU) / University of Tokyo, the Korean Participation Group, Lawrence Berkeley National Laboratory, Leibniz Institut f\"ur Astrophysik Potsdam (AIP),  
Max-Planck-Institut f\"ur Astronomie (MPIA Heidelberg), 
Max-Planck-Institut f\"ur Astrophysik (MPA Garching), 
Max-Planck-Institut f\"ur Extraterrestrische Physik (MPE), 
National Astronomical Observatories of China, New Mexico State University, 
New York University, University of Notre Dame, 
Observat\'ario Nacional / MCTI, The Ohio State University, 
Pennsylvania State University, Shanghai Astronomical Observatory, 
United Kingdom Participation Group,
Universidad Nacional Aut\'onoma de M\'exico, University of Arizona, 
University of Colorado Boulder, University of Oxford, University of Portsmouth, 
University of Utah, University of Virginia, University of Washington, University of Wisconsin, 
Vanderbilt University, and Yale University.

The PS1 has been made possible through contributions by the Institute for Astronomy, the University of Hawaii, the Pan-STARRS Project Office, the Max-Planck Society and its participating institutes, the Max Planck Institute for Astronomy, Heidelberg and the Max Planck Institute for Extraterrestrial Physics, Garching, The Johns Hopkins University, Durham University, the University of Edinburgh, Queen's University Belfast, the Harvard-Smithsonian Center for Astrophysics, the Las Cumbres Observatory Global Telescope Network Incorporated, the National Central University of Taiwan, the Space Telescope Science Institute, the National Aeronautics and Space Administration under Grant No. NNX08AR22G issued through the Planetary Science Division of the NASA Science Mission Directorate, the National Science Foundation under Grant No. AST-1238877, the University of Maryland, and Eotvos Lorand University (ELTE). 

\vspace{5mm}
\facilities{Sloan, PS1, CFHT, Bok}

\bibliography{/Users/yshen/Research/refs}

\end{document}